%% file: main.tex
\documentclass{amsart}
\usepackage[pagewise]{lineno}\linenumbers
\usepackage[utf8]{inputenc}
\usepackage{amsthm,amsmath,amsfonts,amssymb}
\usepackage{soul,color}
\usepackage{longtable}
\usepackage{xcolor}
\usepackage{tabularx}
\usepackage{longtable}
\usepackage{booktabs}
\usepackage{mathtools}
\usepackage{mathabx}
\newcommand{\Fq}{\mathbb{F}_q}
\newtheorem{theorem}{Theorem}[section]

\newtheorem{lemma}[theorem]{Lemma}

\usepackage{color,soul,xcolor}
\usepackage{multicol}
\usepackage{enumitem}
\usepackage{hyperref}
\usepackage{float}
\usepackage[ruled,vlined]{algorithm2e}
\usepackage{enumitem}
\usepackage{comment}
\theoremstyle{definition}
\newtheorem{definition}{Definition}[section]

\title{A Generalization of Cyclic Code Equivalence Algorithm to Constacyclic Codes}
\author{Dev Akre, Nuh Aydin, Matthew J. Harrington, Saurav Pandey}
\date{\today}

\begin{document}
\nolinenumbers

\maketitle
\begin{abstract}
Recently, a new algorithm to test equivalence of two cyclic codes has been introduced which is efficient and produced useful results. In this work, we generalize this algorithm to  constacyclic codes. As an application of the algorithm we found many constacyclic codes with good parameters and properties. In particular, we found 23 new codes that improve the minimum distances of BKLCs.
\end{abstract}


\section{Introduction and Motivation}
A linear code $C$ over a finite field $\Fq$ is a vector subspace of $\mathbb{F}_q^n$ and it has three fundamental parameters: its length($n$), dimension($k$), and minimum distance($d$), and such a code is referred to as a $[n,k,d]_q$ code. One most of the most important problems in coding theory is the optimization of the minimum distance of a linear code. That is, for a given $n$ and $k$, we seek the highest possible $d$. There exist theoretical upper bounds on $d$. A code attaining the upper bound for minimum distance is called (distance) optimal. It should be noted that the upper bounds may actually be unattainable.  One objective in coding theory is to find codes whose minimum distances get as close to the optimal distance as possible. These are called BKLCs (best known linear codes). The online database \cite{database} gives information about BKLCs over small finite fields $\Fq$, $q \leq 9$ up to certain lengths, including lower and upper bounds on $d$. The upper bounds are theoretical, and lower bounds are obtained by explicit constructions. In general, optimal codes are known when either $k$ or $n - k$ is small. This is because calculating the minimum distance is computationally intractable \cite{NPhard}, and the number of linear codes is very large. Consequently, exhaustive searches are not feasible. Hence, we focus on special classes of codes that are promising.

In this paper, we focus on  constacyclic (CC) codes which are a generalization of cyclic codes. Cyclic codes have a prominent place in coding theory for both theoretical and practical reasons. They provide a fundamental link between coding theory and algebra. Both cyclic and CC codes  are used as building blocks in various search algorithms, particularly the ASR search algorithm, which have produced numerous record-breaking quasi-cyclic (QC) \cite{qcgood}, quasi-twisted (QT) \cite{ASR,qtgood,qtgood2, constaStruct}, and multi-twisted (MT) codes \cite{mt, mtgood}. They are even used to generate certain types of quantum codes \cite{quantgood,quantgood2}. 

All of these algorithms benefit from having CC codes with high minimum weights. In a comprehensive implementation of the ASR algorithm, we start with examining all cyclic or CC codes of a given length. Since  equivalent codes have the same parameters, it is redundant to use CC codes that are equivalent to each other. Testing the equivalence between two arbitrary linear codes is computationally expensive. However, an efficient algorithm that is specifically designed for cyclic codes based on cyclotomic cosets has been recently presented \cite{cycliceq}. In this work, we generalize that algorithm to CC codes. Like the cyclic case, the resulting algorithm is faster than the general purpose equivalence test algorithm that is available in computer algebra systems like Magma. This enables us to conduct more extensive searches on  QC, QT and MT codes.

We ran an exhaustive searches for CC codes up to a certain dimension for all finite fields of size $\leq 9$. We obtained a large number of CC codes with better parameters than the best known QT codes from the database \cite{qcdatabase}. A significant number of these codes have  additional desirable properties such as reversibility,  self-orthogonality, and having linear complementary dual (LCD). Furthermore, we found a new code over $GF(7)$ with minimum distance 3 units higher than the current  BKLC  and obtained 21 additional new codes using the standard constructions on it. Another  constacyclic code over $GF(5)$ produced a new code using construction X.

\section{Basic Definitions}
We begin by defining constacyclic (CC) codes. Note that we will be using the usual convention of representing the codewords or vectors in $\Fq^n$ as polynomials in $\mathbb{F}_q[x]$:
$$\Vec{c} = (c_0,c_1,\ldots,c_{n-1}) \in \Fq^n \leftrightarrow
 c(x) = c_0 + c_1 x + \cdots + c_{n-1} x^{n-1} \in \mathbb{F}_q[x]$$
\begin{definition}
A linear code $C$ over  $\Fq$ that is closed under a constacyclic shift $\pi_a$ by a nonzero element $a \in \mathbb{F}_q$ is called a constacyclic code, that is, for any $c = (c_0, c_1,\ldots, c_{n-1}) \in C$,  $\pi_a(c) = (a\cdot c_{n-1}, c_0, c_1,\ldots, c_{n-2}) \in C$ as well.
\end{definition} 

The CC shift of a codeword $c(x)$ corresponds to $x\cdot c(x) \mod x^n-a$. It follows that a CC code is an ideal in the quotient ring $\Fq[x]/\langle x^n-a \rangle$ which is a principal ideal ring. For each CC code $C$, there exists a unique monic generator polynomial $g(x) \in \mathbb{F}_q[x]$ of least degree such that $\langle g(x) \rangle = C$. Hence, $x^n-a=g(x)h(x)$ and there is a one-to-one correspondence between CC codes of length $n$ with shift constant $a$ over $\Fq$ and divisors of $x^n-a$ over $\Fq$. The polynomial $h(x)$ is called the check polynomial of $C$. A CC code is uniquely determined by either the generator polynomial or the check polynomial. For the special case when the shift constant $a$ is $1$, we obtain a cyclic code. Thus, CC codes are generalizations of cyclic codes. CC codes are in turn are a special case of QT codes. 

Two linear codes are called equivalent if one can be obtained from the other by any combination of the following transformations.

\begin{enumerate}
    \item A permutation of the coordinates.
    \item Multiplication of elements in a fixed position by a non-zero scalar in $\Fq$.
    \item Applying a field automorphism $\sigma: \Fq \to \Fq$ to each component of a vector.
\end{enumerate}
If only (1) is used, then the codes are called permutation equivalent. This is a very important
special case since it arises most commonly. Moreover, for  binary codes it is the only form of equivalence.  We can summarize all of these conditions in the following way.

\begin{definition} \cite{equiDef}
Two linear codes $C_1,C_2 \subseteq F^n_q$ are equivalent if there exists a monomial matrix $M$ and an automorphism $\delta$ over $\Fq$ such that $C_1 = C_2 M \delta$. 
\end{definition}

Equivalent codes have identical parameters. Checking for equivalence with existing codes before calculating the minimum distance of a new code might save computational time provided this check is fast enough. There exists a polynomial time reduction from the graph isomorphism problem to code equivalence and thus equivalence checking is not NP-Complete \cite{codeEquiEasy}. However, in reality these checks take very long, especially for large fields with codes of large lengths. We aim to use techniques that are much faster that are specifically  for CC codes.

Through our exhaustive searches, we found many codes that are as good as the current BKLCs and  they additional desirable properties. We define these properties here. For any linear code $C$, its dual code is defined as $C^\perp=\{ v\in \Fq^n: v\cdot c=0 \text{ for all } c\in C \}$ where $v\cdot c$ is the standard inner product in $\Fq^n$. If the dimension of $C$ is $k$, then the dimension of $C^\perp$ is $n-k$. A code $C$ is self-orthogonal if  $C \subseteq C^\perp$, i.e., for any two codewords $a, b \in C$, $a \cdot b = 0$. An $[n,k]_q$ code $C$ is self-dual if  $C = C^\perp$. Note that in this case, the dimensions of $C$ and $C^\perp$ need to be equal. Thus,  $k = n/2$. A code $C$ is dual-containing if $ C^\perp \subseteq C$. All of these properties of codes have been used extensively to find optimal quantum error-correcting codes \cite{self-orthogonal, self-orthogonal2, dual-containing}.

A $[n,k]_q$ code $C$ is linear complementary dual (LCD) if  $C \cap C^\perp = \{ 0 \}$. They were first introduced by Massey \cite{LCD}, and were seen to have an optimal solution for a two-user binary adder channel as well as decoding algorithms that are less complex than that for general linear codes. They are also useful in cryptography by protecting the information managed by sensitive devices, particularly against fault invasive attacks and side-channel attacks (SCA) \cite{LCDuse}.

A code $C$ is reversible if  for any codeword $(c_0,c_1,...,c_{n-2},c_{n-1}) \in C$ then the codeword reversed as $(c_{n-1},c_{n-2},...,c_1,c_0) \in C$. Reversible codes are rate and are useful in cases where the code might be read from any direction \cite{reversible}. 

 Let $C$ be a linear code and  $w(c)$ denote the Hamming weight of codeword $c\in C$. If $w$ takes at most two distinct nonzero values, we call $C$ a two-weight code. These codes have important applications in secret-sharing schemes, and are mathematically related to strongly related graphs \cite{2wgt}.

\section{On equivalence of Constacyclic Codes}

Cyclotomic cosets are useful in the study of cyclic codes in many ways. Some important types of cyclic codes such as BCH codes are defined based on  cyclotomic cosets. Recently, sufficient conditions for two cyclic codes to be equivalent are obtained from cyclotomic cosets \cite{cycliceqAlg,cycliceq,constaStruct}. We will generalize some of these results to CC codes. 

\begin{definition}\cite{ASR}
Let $\gcd(n, q) = 1$. For any $i \in \mathbb{Z}_n$, the $q$-cyclotomic coset of $n$ containing $i$ is the set $S_i = \{iq^j \mod n : j \in \mathbb{N}\}$.
\end{definition}

It is well known that in the case $\gcd(n, q) = 1$ (simple root cyclic codes), there is a one-to-one correspondence between cyclotomic cosets $\mod n$ and irreducible divisors of $x^n-1$. Moreover, following results about equivalence of cyclic codes are obtained based on cyclotomic cosets.  

\begin{theorem} \cite{cycliceq}
\label{theorem1}
Let $g_1(x)$ and $g_2(x)$ be the standard generators of cyclic codes of length n over $\Fq$ and assume $gcd(e, n) = 1$. Then the isometry $\phi : \mathbb{F}_q[x]/\langle x^n - 1\rangle \mapsto \mathbb{F}_q [x]/ \langle x^n - 1\rangle$ given by, 
$$x \mapsto x^e \mod (x^n - 1)$$ 
has the property $g_2(x) = \phi(g_1(x)))$ if and only if the map $\phi : S_{g_1} \mapsto S_{g_2}$ given by $\phi(z) = e^{-1}z \mod n$, where $e^{-1}$ is the multiplicative inverse of $e \mod n$, is a bijection.
\end{theorem}

\begin{theorem} \cite{cycliceq}
\label{theorem2}
Let $g_1(x)$ be the standard generator of a cyclic code of length $n$ over $\Fq$ where $gcd(n, q) = 1$, and let $\delta = \alpha^{-b}$ where $\alpha$ is a primitive $n$th root of unity, such that $n$ divides $b\cdot deg(g_1(x)) \cdot (q - 1)$. Let $K$ be an extension field of $\Fq$ that contains $\delta$. Then the isometry $\phi : K[x]/\langle x^n - 1\rangle \mapsto K[x]/\langle x^n - 1 \rangle$ defined by $\phi( f(x)) = f (\delta x) \mod (x^n - 1)$ has the property that $\phi(g_1(x)) \in \mathbb{F}_q [x]$ and generates a cyclic code of length $n$ over $\Fq$ if and only if the map $\phi : \mathbb{Z}_n 	\mapsto \mathbb{Z}_n$ defined by $\phi(z) = z + b \mod n$ is a bijection such that $\phi(Sg_1 ) = S\phi(g_1)$.
\end{theorem}

In a recent work (\cite{cycliceqAlg}), this correspondence is extended to the repeated root case by considering multisets. Suppose $\gcd(n,q)\not =1$, where $q$ is a power of $p$. We first write $n=n'p^t$ such that $\gcd(p,n')=1$. Next we find the cyclotomic cosets $\mod n'$. Define a function $P$ which takes  cyclotomic cosets to polynomials. Let $\alpha$ be an $n'^{th}$ root of unity, and $S$ be a cyclotomic coset $\mod n'$. We define  $\displaystyle{P(S)=\prod_{i \in S}(x-\alpha^i)}$. Then we use a multiset to describe unions where if an irreducible factor of $x^n-1$ appears multiple times (say $m$ times) in a divisor, then the elements of the cyclotomic coset that corresponds to that divisor appears $m$ times in the multiset.
Hence, a multiset $MS$  is a union of not necessarily distinct cyclotomic cosets $S_1,S_2,...,S_k$ and it corresponds to the polynomial $P(MS)=P(S_1)\cdot P(S_2)\cdots P(S_k).$ Based on this approach, an algorithm to test equivalence of cyclic codes is given in \cite{cycliceqAlg}.


We now generalize these results to CC codes. The following observation is very useful for us to be able to generalize them for constacyclic codes.

Consider the polynomial $x^n-a$ over $\mathbb{F}_q$, where $p$ is the characteristic of $\mathbb{F}_q$.  Given $n$, write  $n = p^{t} n'$ such that $p$ does not divide $n'$. Then, since the map $x\to x^{p^t}$ is a bijection (even an automorphism) on $\mathbb{F}_q$, there exists $b \in \mathbb{F}_q$ such that $a=b^{p^t}$, hence we can write $x^n-a=(x^{n'}-b)^{p^t}$.

\begin{lemma}\label{lemma1}Given $a$ and $b$ as above, we have $|a|=|b|$, where $|\theta|$ denotes the order of $\theta$ in the multiplicative group $\mathbb{F}_q^{*}$.
\end{lemma} 

\noindent \begin{bf}Proof:\end{bf} Let $\alpha$ be a primitive element of $\mathbb{F}_q$. Then for some integer $j$, $b = \alpha^j$, and subsequently $a = \alpha^{jp^t}$. Thus we want to show that $|\alpha^j| = |\alpha^{jp^t}|$. The proof is based on the following well-known theorem from group theory. For a finite cyclic group generated by $g$, the order of a power of $g$ is given by 
$$|g^i| = \frac{|g|}{\gcd(|g|,i)}$$ 

    \noindent From this and the fact that $|\alpha| = q-1$ it follows that 
$$|\alpha^j| = \frac{q-1}{\gcd(q-1,j)}$$ 
and 
$$|\alpha^{jp^t}| = \frac{q-1}{\gcd(q-1,jp^t)}$$ 
As $p$ is the characteristic of $\mathbb{F}_q$, we have $q=p^m$ for some positive integer $m$, and therefore $p^t$ is relatively prime to $q-1$. Hence $\gcd(q-1,j) = \gcd(q-1,jp^t)$, and $|\alpha^j| = |\alpha^{jp^t}|$. Thus, $|a| = |b|$.

\vspace{-2 cm}

\begin{theorem}
\label{theorem3}
Let $g_1(x), g_2(x)$ be generators of constacyclic codes of length $n$ over $\Fq$ with shift constant $a$ (hence $g_1(x)$, $g_2(x)$ are divisors of $x^n - a$). If there is a bijection  $m$ of the form $m(x) = ex + b$ between cyclotomic cosets $\mod nr$ corresponding to $g_1(x)$ and $g_2(x)$, then the constacyclic codes $\langle g_1(x) \rangle$ and $\langle g_2(x) \rangle$ are equivalent.
\end{theorem}

\vspace{-1 cm}

\section{The generalized algorithm}
\label{section:algo}
We now describe our approach in developing an algorithm for checking equivalence based on the theory discussed in the last section. We first obtain $r=Ord_q(a),p=char(\Fq)$ and $n'$ such that $n=n'\cdot p^t$. After finding an $n'\cdot r$th root of unity ($\delta$), for $i=0,1,\ldots,n'-1$ we form cyclotomic cosets $\mod n'r$ of the exponents of $\delta$ of the form $1+i\cdot r$. We then take unions of multi-sets of elements of these cosets where the multiplicity of an element is between 0 and $p^t$. Each multiset corresponds to a polynomial. 


Checking for a linear map is computationally expensive. The most straight-forward approach is to try all values of $a,b \in \{0,\ldots,n'\}$ such that $\gcd(a,n)=1$ and for each $x \in C_1$, $ax + b \in C_2$. If such pair of values exist then the codes are equivalent. The complexity of this process is $O(n^3)$. We save a lot of time by checking if the sum of multiplicities of the elements as well as the distribution of their occurrences are equal before starting to check for a linear map.

Another matter of note is that the choice of the root of unity $\delta$ affects the code that will be stored from an equivalence class. Since different choices for $\delta$ give equivalent codes from the same classes, this is not a problem for our purposes. 

\input{Algorithms/Algorithm1}

\section{Performance and limitations}

The following table compares our CC\_CosetEq Function with Magma's IsEquivalent Function. It can be seen that our method is always faster, considerably so in cases where the codes are actually equivalent. However, Magma's function is more versatile while our method is tailor-made for CC codes. Since Magma has no other version of testing code equivalence available, we will make the comparison between our algorithm and Magma's algorithm. In the tables below, a polynomial is represented as a list containing only coefficients in order to save space. The ordering is such that coefficients of lowest degree term is in the left-most position. For instance, the polynomial $1\cdot x+2\cdot x^2+ 3\cdot x^3$ will be represented as $[0123]$. The online Magma calculator \cite{magma} is used for the comparison. It has a time limit of 120 seconds and memory limit of about 360 MB. The entries in the table with "DNF" refers to the programs that did not finish either due to the online calculator's time or memory limit. The Magma IsEquivalent Function also only works for small prime fields or fields of size less than or equal to $4$.
\input{Tables/Perforx}

It is important to note that the algorithm only checks for a sufficient condition of equivalence. This means that the function might return False even if the codes are actually equivalent. For instance, consider the constacyclic codes with $n=32, a=1$ with generators $g_1= 222120111202021, g_2=
222112021022021$. Our algorithm does not detect equivalence between these codes even though they actually are equivalent. However, computational evidence seems to suggest that this is a rare occurrence. The next section about partitioning CC codes into equivalent classes furthermore shows that in many cases  the number of codes that need to be searched is reduced by a large amount. We can also check for equivalence in $GF(8)$ and $GF(9)$, which was not possible with Magma's function. Thus, for the goal of executing an exhaustive search on CC  codes, we find our method more viable even with the potential for rare occurrence of redundancies.

\section{Applications of the algorithm}

This is an application of the algorithm given in section 4. We partition constacyclic codes of given field, length and shift constant into equivalence classes. It is purely based on cyclotomic cosets and using their combinations. We break the elements list into the component cyclotomic cosets $\mod n'\cdot r$. Then we take unions of not necessarily distinct cosets up to $p^t$ times. Using the new algorithm we check if the codes generated by this union of cosets is equivalent to any previously seen code. In the end, we convert the multisets to generator polynomials using the map $P$ we defined in section 3 and store them. 

\input{Algorithms/Algorithm2}

Table \ref{table:2} shows the effectiveness of our constacyclic partition algorithm for some sample code lengths. Here, $q$ is the size of the finite field, $n$ represents the length of the code, $a$ represents the shift constant, \emph{total} represents the total number of divisors of $x^n - a$, \emph{new} represents the number of polynomials generated by our algorithm, and \emph{net} represents the difference between \emph{total} and \emph{new}. Here, \emph{net} is the reduction in the number of codes due to code equivalence and thus is a good indication of a possible benefit of our algorithm when considering a code of a given length. The final column, \emph{Percent decrease}, shows the percentage of reduction in the total number of codes caused by our algorithm. This value is simply the ratio of \emph{net} to \emph{total} multiplied by 100.
\input{Tables/PerformanceTable}

\newpage

\section{Results}

This section contains our findings from the partition algorithm. Tables \ref{tab:prop_table1} - \ref{tab:prop_table10} below show CC codes obtained from our searches using the new algorithm. These codes are as good as the current BKLCs \cite{database}, and better than currently known QT Codes \cite{qcdatabase}. Furthermore, the ones listed here have additional properties by which they are classified into tables. For brevity's sake, we only list some of all (638) such codes we obtained. 

The first column specifies the parameters of the code, the second lists the shift constant, and the third column gives either the generator $g$ or the parity check polynomial $h$, whichever is more concise.  

\input{Tables/Properties}

We found 453 more codes  that are better than best known QT codes and as good as current BKLCs without any additional properties. However, they are very simple to construct. Many of the BKLCs with the same parameters as our codes have complicated constructions involving multiple steps. For instance, consider the BKLC $[68,52,8]_5$ from \url{http://www.codetables.de} \cite{database} is constructed in 7 seven steps.
Our construction for a code just as good is just one step- CC code with length $68$, shift constant $2$ and generator $11040231132244121$. This construction is less complicated and the code is easier to replicate. Thus, our codes are better alternatives than the ones listed in the database with the same parameters. Additionally, we have found a total of 23 new linear codes with higher minimum distances than the currentky BKLCs listed in \cite{database}.

We  found a new $[65,51,8]_5$ code from our search results using construction X. This code is better than currently known linear codes and can be constructed as follows:

\begin{enumerate}
    \item [1]:  [63, 50, 6] Constacyclic Code over GF(5)\\
     $a = 1$, $g = 1133013103311$
     \item [2]:  [63, 51, 8] Constacyclic Code over GF(5)\\
     $a = 1$, $g = 40303432120201$
     \item [3]: [2, 1, 2] Cyclic Linear Code over GF(5)\\
     RepetitionCode of length 2
     \item [4]: [65,51,8] Linear Code over GF(5)\\
     Construct X using [1], [2] and [3]
\end{enumerate}

We also found a $[93,15,58]_7$ code whose minimum distance is 3 units larger than the current BKLC having same length and dimension.  \begin{enumerate}
  \item [] [93,15,58] Constacyclic Code over GF(7)\\
     $a = 2$,  $g= 4340263542221420141536623563456464141141502150610214634\\203012246201352136540611$
\end{enumerate}

\section{Recursive Standard Constructions}

In the course of the search, we found codes that beat the currently best known minimum distance by more than one unit. In the case of the $[93,15,58]_7$ code, its minimum distance was 3 higher than the BKLC that preceded it. This means that there was high potential for  other codes derived from this code to produce additional record breakers by using such standard constructions as extension, puncturing, and shortening. Each of these constructions are implemented in Magma \cite{magma}, which allowed us to implement the following algorithm:



\begin{algorithm}[H]
\SetAlgoLined

\textbf{Input:} C: A good code with parameters $[n,k,d]_q$\;
\textbf{Input:} ShortenLimit: A constant that will determine how many places we can shorten at once\;
  \SetKwProg{Fn}{Function}{:}{}
  \Fn{RecursivelyModify(C,foundparams)}{
  \Fn{Check(C,foundparams)}{
      \If{Cprime is better than the corresponding BKLC and Parameters(Cprime) not in foundparams}{
        Print(Cprime)\;
        \textbf{return}
        Concatenate(foundparams,RecursivelyModify(Cprime,foundparams);
      }
      \textbf{return} [ ]\;
  }
  foundparams = Append(foundparams,Parameters(C))\;
  Cprime=ExtendCode(C)\;
  foundparams+= Check(Cprime,foundparams)\;
  
  
  CprimeP=CprimeS= [1,1,1] trivial code\;
  \For{s from 1 to n}{
  CtempP=PunctureCode(C,i)
  CtempS=ShortenCode(C,s)\;
  \If{CtempP is better than CprimeP}
  {CprimeP=CtempP\;}
  \If{CtempS is better than CprimeS}
  {CprimeS=CtempS\;}
  }
  
  foundparams+= Check(CprimeP,foundparams)\;
  foundparams+= Check(CprimeS,foundparams)\;
  \textbf{return} foundparams\;
 }
 RecursivelyModify(C,[ ]);
 \caption{Recursive Code Modification}
\end{algorithm}


Through the use of this algorithm, we found 21 new codes stemming from the $[93,15,58]_7$ code, $C_1$. Any code derived from  $C_1$ or its derivative by extension, puncturing or shortening is the name of original code appended by 'e','p' or 's' respectively. For instance C1ees is C1 extended twice and then shortened once. Puncturing and shortening is done from the best position possible. 

\input{Tables/GreatCodesGF7}

This algorithm is especially useful for producing new codes from a good code which beats the corresponding minimum distance record by more than 1 unit. A Magma file to execute it can be obtained by contacting the authors of this paper.

\input{bibliography}
\end{document}

%% file: Algorithms/Algorithm1.tex
\begin{algorithm}[H]
\SetAlgoLined
\KwIn{ F \CommentSty{(Finite Field of size $q$)}, n \CommentSty{(Length)}, a \CommentSty{(Shift Constant)} , $g_1$ and $g_2$ \CommentSty{(generator polynomials of $C_1$ and $C_2$)}\;}
\KwOut{ True \CommentSty{(if algorithm detects equivalence)}, False \CommentSty{(otherwise)}\;}
 
\SetKwFunction{FMain}{CC\_CosetEq}
\SetKwProg{Fn}{Function}{:}{}
\Fn{\FMain{$F,n,a,g_1,g_2$}}{
 $r = Order(a)$\;
 $p = Characteristic(F)$\;
 $n'$ such that $n = n'\cdot (p)^t$ for highest possible $t \in \mathbb{N}$\;
 $EF = ExtensionField(F)$ defined by Irreducible Polynomial in $F$ of Degree $(Order(n'\cdot r \mod q))$\;
 elements = [$1+i\cdot r$: $i$ from $0$ to $n'-1$]\;
 rou = $(n'r)^{th}$ root of unity in $EF$\;
 
 \For{$i$ in elements}{
 coset1[i] is the largest integer $y$ such that $(x-rou^i)^y|g_1$\;
 coset2[i] is the largest integer $y$ such that $(x-rou^i)^y|g_2$\;
 }
 Equivalent = false\;
 \If{Sum(coset1) == Sum(coset2)}{
   \If{Distribution(coset1) == Distribution(coset2)}{
     \If{existsLinearMap(coset1,coset2)}{
       Equivalent = true\;
     }
   }
 }
 \textbf{return} Equivalent\ ;
 }
\textbf{End Function}\;
 \caption{Algorithm employing the constacyclic coset equivalence checks to decide equivalence between two constacyclic codes}
\end{algorithm}

%% file: Tables/Perforx.tex
\begin{table}[H]
    \centering
    
    \caption{Performance Comparison of CosetEq method vs inbuilt IsEquivalent function}
    
    \tabcolsep=0.11cm
    \begin{tabular}{l|l|l|l|l|l|l|l|l}
    \multicolumn{5}{l|}{} &
    \multicolumn{2}{l|}{\bf CC\_CosetEq} &
    \multicolumn{2}{l}{\bf IsEquivalent}\\

    q & n & g(s) & a & equiv & \begin{tabular}{l}
    \footnotesize
      CPU \\ 
      \footnotesize
      time(s)
    \end{tabular}  & \begin{tabular}{l}
    \footnotesize
         Memory\\\footnotesize
         (MB)
    \end{tabular} &\begin{tabular}{l}
      \footnotesize CPU \\ \footnotesize
      time(s)
    \end{tabular}&
    \begin{tabular}{l}
        \footnotesize Memory\\\footnotesize
         (MB)
    \end{tabular}\\ 
    \hline&&&&&&&\\
    $2$&$210$&
    \begin{tabular}{l}
      $[ 1 1 1 0 0 1 1 0 0 1 1 0 0 1 0 0 1 ]$ \\ 
      $[ 1 1 0 0 0 1 1 0 0 1 1 0 1 0 1 0 1 ]$
    \end{tabular} & 1& True & 0.120 & 32 & 101.300 & 32\\
    
    $3$& $90$ &
    \begin{tabular}{l}
         $[12011 ]$  \\
         $[11021 ]$ 
    \end{tabular}&2 &False &0.000 &32 & \multicolumn{2}{l}{\footnotesize DNF- Memory limit} \\
    
    $4$& $60$ &
    \begin{tabular}{l}
         $[ A001001]$  \\
         $[ A00A001]$ 
    \end{tabular}&$A$ & True& 0.000& 32 & 0.700 & 32 \\

    $ 5$& $ 68$ &
    \begin{tabular}{l}
         $[ 11434131132402021]$  \\
         $[ 14424434122103031]$ 
    \end{tabular}&3 &True & 0.020 & 32& \multicolumn{2}{l}{\footnotesize DNF- Time limit} \\
    
        $ 7$& $ 53$ &
    \begin{tabular}{l}\tiny
         $[430340635506060303635046031 ]$  \\
         \tiny
         $[ 460150604034306230205052061]$ 
    \end{tabular}&4 &False &0.020 &32& \multicolumn{2}{l}{\footnotesize DNF- Time limit} \\

    \end{tabular}
    \label{perf_table}
\end{table}

%% file: Algorithms/Algorithm2.tex
\begin{algorithm}[htbp!]
\SetAlgoLined
\KwIn{ q \CommentSty{(size of finite field)}, n \CommentSty{(Length)}, a \CommentSty{(Shift Constant)}\;}
\KwOut{generatorList \CommentSty{(List of unequivalent generators dividing $x^n-a$)}\;}
 $F = FiniteField(q)$\;
 $r = Order(a)$\;
 $p = Characteristic(F)$\;
 $n'$ such that $n = n'\cdot (p)^t$ for highest possible $t \in \mathbb{N}$\;
 $EF = ExtensionField(F)$ defined by Irreducible Polynomial in $F$ of Degree $(Order(n'\cdot r \mod q))$\;
 CycCosets = $[ ]$\;
 elements = [$1+i\cdot r$: $i$ from $0$ to $n'-1$]\;
 \For{$i$ in elements}{
    \If{$i$ not in CycCosets}{
      CycCosets+= $\{iq^j: j = 0,1,\ldots\}$\;
    }
 }
 $numCosets = \#CycCosets$\;
 $rou = (n'r)^{th}$ root of unity in $EF$\;
 $totalnum = (p^t+1)\wedge numCosets - 2$; \CommentSty{(Non-trivial divisors of $x^n-a$)}\\
 UneqCosets = [ ]\;
 generatorList = [ ]\;
 \For{i from 1 to totalnum}{
    powers = $i$ base ($p^t + 1$)\;
    TempCoset = $\{ \}$; \CommentSty{(Multi-Set)}\\
    \For{j from 1 to numCosets}{
        TempCoset += CycCosets$[j]\wedge (j^{th}Digit(i))$
    }
    Equivalent = false\;
    \For{CheckCoset in UneqCosets}{
         \If{Sum(CheckCoset) == Sum(TempCoset)}{
           \If{Distribution(CheckCoset) == Distribution(TempCoset)}{
             \If{existsLinearMap(CheckCoset,TempCoset)}{
               Equivalent = true\;
     }
   }
 }
    }
    \If{Equivalent == true}{\
        UneqCosets+= TempCoset\;
        generator = 1\;
        \For{j in TempCoset}{
        generator*= $(x-rou^j)\wedge (Multiplicity(j))$\;
        }
        generatorList+= generator\;
    }
 }
 Print(generatorList)\;

 \caption{Algorithm that returns list of unequivalent generators for CC codes of length $n$, shift constant $a$}
\end{algorithm}

%% file: Tables/PerformanceTable.tex
\begin{table}[hbt!]
\centering
\caption{Reduction in the number of codes from our algorithm.}
\begin{tabular}{||c c c c c c c||} 
 \hline
 q & n & a & total & new & net &Percent decrease \\ [0.5ex] 
 \hline\hline
2 & 93 & 1 & 16382 & 2798 & 13584 & 82.92 \\
2 & 105 & 1 & 32766 & 9598 & 23168 & 70.71 \\
2 & 120 & 1 & 59047 & 32803 & 26244 & 44.45 \\
2 & 124 & 1 & 78123 & 13173 & 64950 & 83.14 \\
3 & 146 & 2 & 8190 & 536 & 7654 & 93.46 \\
3 & 122 & 2 & 8190 & 455 & 7735 & 94.44 \\
3 & 130 & 2 & 32766 & 969 & 31797 & 97.04 \\
5 & 124 & 2 & 2046 & 26 & 2020 & 98.73 \\
5 & 90 & 2 & 7774 & 3074 & 4700 & 60.46 \\
5 & 52 & 2 & 8190 & 1380 & 6810 & 83.15 \\
5 & 104 & 2 & 8190 & 469 & 7721 & 94.27 \\
5 & 52 & 4 & 16382 & 2129 & 14253 & 87.00 \\
5 & 108 & 4 & 16382 & 1269 & 15113 & 92.25 \\
5 & 60 & 4 & 46654 & 12839 & 33815 & 72.48 \\
5 & 120 & 4 & 46654 & 696 & 45958 & 98.51 \\
7 & 76 & 6 & 16382 & 1126 & 15256 & 93.13 \\
7 & 90 & 6 & 32766 & 1519 & 31247 & 95.36 \\
7 & 86 & 6 & 32766 & 655 & 32111 & 98.00 \\ [1ex]
 \hline
\end{tabular}
\smallskip 

\label{table:2}
\end{table}

%% file: Tables/Properties.tex
\begin{longtable}{l|l|l}
\caption{New  CC Codes that are Self-Orthogonal (19 of 122)}\\
        $[n,k,d]_q$ & a & $h$\\
        \hline&\\
$ [91,39,20]_2$ & $1 $ & $1100001011011010010010100011110110111111$\\
$ [79,39,16]_2$ & $1 $ & $1110110000010110101111001111011100011001$\\
$ [223,37,72]_2$ & $1 $ & $11111000101011001111101100111000101101$\\
$ [83,41,21]_3$ & $1 $ & $221200201021221100222200221210011200121201$\\
$ [164,26,72]_3$ & $2 $ & $222120021201020122012210021$\\
$ [82,24,30]_3$ & $2 $ & $1222011211212222111201211$\\
$ [19,9,8]_4$ & $1 $ & $1A^20A^2A^2AA0A1$\\
$ [129,21,64]_4$ & $1 $ & $111A0A^2A^211AAA^2A^211AA0A^2111$\\
$ [38,18,13]_5$ & $1 $ & $4142302342133022111$\\
$ [52,14,25]_5$ & $4 $ & $322133301433401$\\
$ [31,12,14]_5$ & $1 $ & $1403040341241$\\
$ [47,23,17]_7$ & $1 $ & $654323415250330435200061$\\
$ [50,20,20]_7$ & $6 $ & $112364342641233364261$\\
$ [85,16,48]_7$ & $1 $ & $12105062226642241$\\
$ [79,13,49]_8$ & $1 $ & $1A^41A^4A^5A^6A^300A^6AA^5A^61$\\
$ [19,9,10]_9$ & $1 $ & $2A^52A^2A^3A^5A^21A^31$\\
$ [31,15,12]_9$ & $1 $ & $2A^5A^3A^7A^32A^6A^7AA^61A^5AA^5A^31$\\
$ [37,18,14]_9$ & $1 $ & $1A^310A^6A^5A12221A^3A^7A^201A1$
    \label{tab:prop_table1}
\end{longtable}

\begin{longtable}{l|l|l}
    \caption{New  CC Codes that are Dual Containing (16 of 139)}\\
        $[n,k,d]_q$ & a & $g$\\
        \hline&\\
$ [133,112,6]_2$ & $1 $ & $1110101111110000110001$\\
$ [151,106,13]_2$ & $1 $ & $1010100111001100110111000110110101001010111001$\\
$ [93,48,14]_2$ & $1 $ & $1001111001101000011101000000011001000010110001$\\
$ [109,82,10]_3$ & $1 $ & $2220110212021200001101101101$\\
$ [82,58,10]_3$ & $2 $ & $1100210002021101000220021$\\
$ [133,112,8]_4$ & $1 $ & $1A^20A^2A^2A^2A^2AAA^2A^2110AA001AA^21$\\
$ [71,51,10]_5$ & $1 $ & $103402021440032402131$\\
$ [52,34,10]_5$ & $4 $ & $3103324404332410421$\\
$ [44,23,12]_5$ & $1 $ & $2044142410012132403401$\\
$ [58,44,8]_7$ & $1 $ & $650012241422041$\\
$ [40,28,8]_7$ & $6 $ & $1060426323511$\\
$ [47,24,16]_7$ & $1 $ & $610005243044025263454321$\\
$ [79,66,8]_8$ & $1 $ & $1A^6A^5AA^600A^3A^6A^5A^41A^41$\\
$ [37,28,7]_9$ & $1 $ & $2A^61A^6AA^7A^62A^61$\\
$ [31,16,11]_9$ & $1 $ & $2A^7AA^5A2A^2A^5A^3A^21A^7A^3A^7A1$\\
$ [37,19,13]_9$ & $1 $ & $1A10A^2A^7A^312221AA^5A^601A^31$
    \label{tab:prop_table2}
\end{longtable}

\begin{longtable}{l|l|l}
    \caption{New  CC Codes that are LCD (12 of 79)}\\
    
        $[n,k,d]_q$ & a & $g$ or $h$\\
        \hline&\\
$ [57,30,14]_4$ & $A $ & $111AA^2A^2AAA^2A11111111A^2AA^2A^2AAA^2111$\\
$ [105,84,8]_4$ & $A $ & $A^2000AA^2A^200A^2A^2A0A^2A01A^2AA01$\\
$ [171,18,96]_4$ & $A $ & $h = A^200AA^201A^20A01A^201A001$\\
$ [68,52,8]_5$ & $2 $ & $11040231132244121$\\
$ [52,24,17]_5$ & $2 $ & $h = 4023014140102413440204101$\\
$ [46,24,13]_5$ & $2 $ & $31443110010104003414241$\\
$ [86,72,8]_7$ & $3 $ & $440151543452041$\\
$ [40,20,14]_7$ & $3 $ & $h = 262441353161263215461$\\
$ [50,16,26]_7$ & $3 $ & $h = 24632222431542661$\\
$ [10,6,5]_9$ & $A $ & $A^2A^7A^2A^61$\\
$ [34,26,6]_9$ & $A $ & $2A^3A^5A^71A^6A^311$\\
$ [58,14,33]_9$ & $A $ & $h = A^7A^6A^501A^6A^7AA^62A^50111$
    \label{tab:prop_table3}
\end{longtable}

\begin{longtable}{l|l|l}
    \caption{New  CC Codes that are Reversible (19 of 29)}\\
    
        $[n,k,d]_q$ & a & $g$ or $h$\\
        \hline&\\
$ [204,191,4]_2$ & $1 $ & $10101100110101$\\
$ [180,166,4]_2$ & $1 $ & $111010101010111$\\
$ [168,154,4]_2$ & $1 $ & $100110000011001$\\
$ [72,61,4]_2$ & $1 $ & $110101101011$\\
$ [30,23,4]_3$ & $1 $ & $11122111$\\
$ [12,7,4]_3$ & $1 $ & $101101$\\
$ [6,3,3]_3$ & $1 $ & $h = 1221$\\
$ [34,20,8]_4$ & $1 $ & $111A^2A01010AA^2111$\\
$ [68,61,4]_4$ & $1 $ & $1A^2A00AA^21$\\
$ [65,58,4]_5$ & $1 $ & $40141401$\\
$ [30,25,4]_5$ & $1 $ & $142241$\\
$ [15,10,4]_5$ & $1 $ & $424131$\\
$ [56,50,4]_7$ & $1 $ & $1124211$\\
$ [56,45,6]_7$ & $1 $ & $166534435661$\\
$ [18,11,6]_8$ & $1 $ & $1A^51A^6A^61A^51$\\
$ [18,14,4]_8$ & $1 $ & $1A^20A^21$\\
$ [36,31,4]_8$ & $1 $ & $1A^6A^2A^2A^61$\\
$ [30,25,4]_9$ & $1 $ & $1A^311A^31$
    \label{tab:prop_table4}
\end{longtable}

\begin{longtable}{l|l|l}
    \caption{New  CC Codes that are Self-Dual}\\
        $[n,k,d]_q$ & a & $h$\\
        \hline&\\
$ [28,14,9]_3$ & $2 $ & $221211000122221$\\
$ [8,4,5]_7$ & $6 $ & $15221$
    \label{tab:prop_table5}
\end{longtable}

\begin{longtable}{l|l|l}
    \caption{New  CC Codes that are Self-Orthogonal and Reversible}\\
        $[n,k,d]_q$ & a & $g$\\
        \hline&\\
$ [10,3,6]_4$ & $1 $ & $(D)1A^2A^21$\\
$ [7,3,5]_7$ & $1 $ & $(D)6341$\\
$ [18,3,14]_8$ & $1 $ & $(D)1A^3A^31$
    \label{tab:prop_table6}
\end{longtable}

\begin{longtable}{l|l|l}
    \caption{\footnotesize{New  CC Codes that are Dual-Containing and Reversible (4 of 9)}}\\
        $[n,k,d]_q$ & a & $g$\\
        \hline&\\
$ [10,7,3]_5$ & $1 $ & $4411$\\
$ [56,52,3]_7$ & $1 $ & $16361$\\
$ [56,51,4]_7$ & $1 $ & $102201$\\
$ [28,23,4]_7$ & $1 $ & $134431$
    \label{tab:prop_table7}
\end{longtable}

\begin{longtable}{l|l|l}
    \caption{New  CC Codes that are LCD and Reversible (20 of 245)}\\
        $[n,k,d]_q$ & a & $g$ or $h$\\
        \hline&\\
$ [171,134,10]_2$ & $1 $ & $10010000001000110111101100010000001001$\\
$ [129,87,13]_2$ & $1 $ & $1011111011001100111011101110011001101111101$\\
$ [65,40,10]_2$ & $1 $ & $10001101101011010110110001$\\
$ [146,122,8]_3$ & $1 $ & $1122121011100011101212211$\\
$ [82,49,14]_3$ & $1 $ & $1211200010200021001200020100021121$\\
$ [74,38,16]_3$ & $2 $ & $1101011222112200022200022112221101011$\\
$ [29,15,11]_4$ & $1 $ & $1A0AA^21A^2AA^21A^2A0A1$\\
$ [65,33,16]_4$ & $1 $ & $1A^2A^20AA01A^2100A^21101011A^2001A^210AA0A^2A^21$\\
$ [241,228,6]_4$ & $1 $ & $1A^21A00AA00A1A^21$\\
$ [67,23,27]_5$ & $1 $ & $h = 421134030211443020124431$\\
$ [67,22,28]_5$ & $1 $ & $h = 14324002204340220042341$\\
$ [41,21,13]_5$ & $1 $ & $100203331020133302001$\\
$ [29,14,12]_5$ & $1 $ & $h = 144224030422441$\\
$ [50,21,20]_7$ & $1 $ & $h = 6515262441166335152621$\\
$ [29,15,11]_7$ & $1 $ & $104516141615401$\\
$ [57,13,33]_8$ & $1 $ & $h = 1A^2A^3A^5A^60A^3A^30A^6A^5A^3A^21$\\
$ [65,52,8]_8$ & $1 $ & $1A^6A^3A^4A^2A^311A^3A^2A^4A^3A^61$\\
$ [29,14,12]_9$ & $1 $ & $h = 1A1A^2A^32A^2A^2A^22A^3A^21A1$\\
$ [41,24,12]_9$ & $1 $ & $2A^6A^7A^5A^6A^6A^70A^2A^60A^3A^2A^2AA^3A^21$\\
$ [73,12,47]_9$ & $1 $ & $h = 1A^50AA^60A^30A^6A0A^51$
    \label{tab:prop_table8}
\end{longtable}

\begin{longtable}{l|l|l}       
    \caption{New  CC Codes that are Self-Orthogonal and two-weight (3 of 6)}\\
        $[n,k,d]_q$ & a & $h$\\
        \hline&\\
$ [7,3,4]_4$ & $1 $ & $1011$\\
$ [22,5,12]_3$ & $1 $ & $102221$\\
$ [12,4,6]_3$ & $2 $ & $11221$
    \label{tab:prop_table9}
\end{longtable}

\begin{longtable}{l|l|l}

    \caption{New  CC Codes that are LCD and two-weight}\\
        $[n,k,d]_q$ & a & $h$\\
        \hline&\\
$ [17,4,12]_4$ & $1 $ & $11A11$\\
$ [26,4,20]_5$ & $2 $ & $41331$
\label{tab:prop_table10}
\end{longtable}

\begin{longtable}{l|l|l}

    \caption{New  CC Codes that are Self-Orthogonal, two-weight and reversible}\\
        $[n,k,d]_q$ & a & $h$\\
        \hline&\\
$ [34,4,24]_4$ & $1 $ & $11A11$\\
$ [10,4,4]_2$ & $1 $ & $11111$
\label{tab:prop_table11}
\end{longtable}

%% file: Tables/GreatCodesGF7.tex
\begin{multicols}{3}
{\renewcommand\labelitemi{}
\begin{itemize}[leftmargin=*]
\item C1: [93,15,58]
\item C1e: [94,15,58]
\item C1ee: [95,15,58]
\item C1ees: [94,14,58]
\item C1eesp: [93,14,58]
\item C1eespp: [92,14,58]
\item C1eespppp: [90,14,56]
\item C1eesppps: [90,13,57] 
\item C1eespps: [91,13,58]
\item C1eesppse: [92,13,58]
\item C1eesps: [92,13,58]
\item C1p: [92,15,57]
\item C1pp: [91,15,56]
\item C1ppp: [90,15,56]
\item C1pppp: [89,15,56]
\item C1ppppp: [88,15,56]
\item C1pppppp: [87,15,56]
\item C1ppppppp: [86,15,56]
\item C1pppps: [88,14,54]
\item C1ppppse: [89,14,54]
\item C1ppss: [89,13,56]
\item C1ppssp: [88,13,55]
\end{itemize}
}
\end{multicols}